\documentclass[pra,twocolumn]{revtex4}%
\usepackage{amsfonts}
\usepackage{amsmath}
\usepackage{amssymb}
\usepackage{subfigure}
\usepackage{graphicx}%
\providecommand{\U}[1]{\protect\rule{.1in}{.1in}}

\begin{document}
\title{Strong photon blockade with intracavity electromagnetically induced
transparency in a blockaded Rydberg ensemble }
\author{G. W. Lin$^{1}$}
\email{gwlin@ecust.edu.cn}
\author{Y. H. Qi$^{1}$}
\author{X. M. Lin$^{2}$}
\author{Y. P. Niu$^{1}$}
\email{niuyp@ecust.edu.cn}
\author{S. Q. Gong$^{1}$}
\email{sqgong@ecust.edu.cn}
\affiliation{$^{1}$Department of Physics, East China University of Science and Technology,
Shanghai 200237, China}
\affiliation{$^{2}$Fujian Provincial Key Laboratory of Quantum Manipulation and New Energy
Materials, Fujian Normal University, Fuzhou 350108, China}

\begin{abstract}
We consider the dynamics of intracavity electromagnetically induced
transparency (EIT) in an ensemble of strongly interacting Rydberg atoms. By
combining the advantage of variable cavity lifetimes with intracavity EIT and
strongly interacting Rydberg dark-state polaritons, we show that such
intracavity EIT system could exhibit very strong photon blockade effect.

\end{abstract}

\maketitle

\section{Introduction}

Photon blockade \cite{111,Imamo}, a nonlinear phenomenon where a single photon
can block the presence of other photons, has attracted significant attention
for its important potential applications in quantum optics and quantum
information science. The photon blockade has been theoretically and
experimentally researched in a variety of systems, such as cavity quantum
electrodynamics \cite{111,Imamo,Werner,Birnbaum,Ridolfo,Majumdar,Liu,1,2}.
However, experimental realization of strong photon blockade is still a
challenging pursuit, because the observation of strong photon blockade
requires large nonlinearities with respect to the decay rate of the system.
Recent theoretical protocols for unconventional photon blockade are proposed
based on weak nonlinearities \cite{3,4,5,6,7,8}, but these protocols are
limited to the small occupancy in the cavity \cite{3,6}.

The effects of electromagnetically induced transparency (EIT) play a pivotal
role in quantum nonlinear optics \cite{Schmidt}. Optical nonlinearities
typically arise from higher-order light-atom interactions, such that the
nonlinearities at the single-photon level is very small. To overcome this
limitation, several groups have been studying nonlinear optics and realizing
quantum-information processing with EIT in cold Rydberg gases
\cite{Pritchard,Ates,Sevin,Petrosyan,J. D.
P,Friedler,He,Shahmoon,Saffman,Mohapatra,aa,bb,cc,dd,ee}. In particular, the
experiments with EIT in strongly interacting Rydberg atoms has demonstrated
for quantum nonlinear absorption filter \cite{Peyronel}, single-photon switch
\cite{n Baur}, and single-photon transistor \cite{Gorniaczyk,Tiarks}.

An EIT medium placed in a cavity, which is known as intracavity EIT termed by
Lukin et al. \cite{Lukin}, can substantially affect the properties of the
resonator system. The intracavity EIT provides an effective way to
significantly enhance the cavity lifetime \cite{Laupr} and narrow the cavity
linewidth \cite{Wang,Hernandez,Wu1}. Recently, it was shown that intracavity
EIT could be used for the optical control of photon blockade and antiblockade
effects with a single three-level atom trapped in a cavity \cite{Souza} and
the realization of quantum controlled-phase-flip gate between a flying photon
qubit and a stationary atomic qubit assisted by Rydberg blockade interaction
\cite{Hao}. In this paper, we consider the dynamics of intracavity EIT in an
ensemble of strongly interacting Rydberg atoms and show that such intracavity
EIT system could exhibit very strong photon blockade effect. A photon from the
probe field is injected to the cavity to form the first Rydberg dark-state
polariton. However, injection of a second photon will be blocked, because of
strong nonlinear coupling\ of two Rydberg dark-state polaritons. By combining
the advantage of variable cavity lifetimes with intracavity EIT and strongly
interacting Rydberg dark-state polaritons, the nonlinearity strength of the
polaritons could be much larger than the decay rates of the system, and thus
very strong photon blockade effect would be observed.

\section{Physical model}

\begin{figure}[ptb]
\includegraphics[width=3.2in]{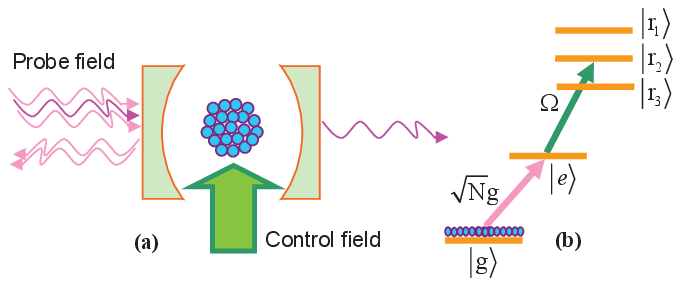}\newline\caption{(Color online) (a)
Schematic setup for the strong photon blockade, by combining the advantage of
variable cavity lifetimes with intracavity EIT and strongly interacting
Rydberg dark-state polaritons. (b) The relevant atomic level structure and
transitions.}%
\end{figure}

As illustrated in Fig.1(a), our model consists of an ensemble of N cold
Rydberg atoms inside an two-sided optical cavity. The concrete atomic level
structure and relevant transitions are shown in Fig.1(b). The atomic
transition $\left\vert g\right\rangle \leftrightarrow\left\vert e\right\rangle
$ is resonantly coupled by the cavity mode with coupling strength $g$, while
the control field with Rabi frequency $\Omega$ resonantly drives the
transition $\left\vert e\right\rangle \leftrightarrow\left\vert r_{2}%
\right\rangle $. Thus they form the three-level EIT configuration, with the
interaction Hamiltonian \cite{Hao} $H_{1}=\sqrt{N}gC_{e}^{\dagger}a+C_{r_{2}%
}^{\dagger}C_{e}\Omega+H.c.$, where $a$ is the annihilation operator of the
cavity mode, $C_{\mu}^{\dagger}=\frac{1}{\sqrt{N}}%
{\displaystyle\sum\nolimits_{j=1}^{N}}
\left\vert \mu\right\rangle _{j}\left\langle g\right\vert $ with $\mu=e$,
$r_{1\text{,}}$ $r_{2\text{,}}$ and $r_{3}$ (the notations $C_{r_{1}}%
^{\dagger}$ and $C_{r_{3}}^{\dagger}$ will appear later) denote the collective
atomic operators, and we have assumed that almost all atoms are in the ground
state, $\left\vert G\right\rangle =\prod_{j=1}^{N}\left\vert g_{j}%
\right\rangle $, at all times. One can define the following polariton
operators $b_{0}^{\dagger}=\cos\theta a^{\dagger}-\sin\theta C_{r_{2}%
}^{\dagger}$ \cite{Fleisch}, $b_{1}^{\dagger}=(\sin\theta a^{\dagger}%
+C_{e}^{\dagger}+\cos\theta C_{r_{2}}^{\dagger})/\sqrt{2}$, $b_{2}^{\dagger
}=(\sin\theta a^{\dagger}-C_{e}^{\dagger}+\cos\theta C_{r_{2}}^{\dagger
})/\sqrt{2}$, which describe the quasiparticles formed by combinations of
photon and atom excitations, here $\cos\theta=\Omega/\sqrt{Ng^{2}+\Omega^{2}}$
and $\sin\theta=\sqrt{N}g/\sqrt{Ng^{2}+\Omega^{2}}$. Then the Hamiltonian
$H_{1}$ with a single photon inputted can be diagonalised and represented by%
\begin{equation}
H_{1}=E_{0}b_{0}^{\dagger}b_{0}+E_{1}b_{1}^{\dagger}b_{1}+E_{2}b_{2}^{\dagger
}b_{2},
\end{equation}
where $E_{0}=0$, $E_{1}=\sqrt{Ng^{2}+\Omega^{2}}$, and $E_{2}=-\sqrt
{Ng^{2}+\Omega^{2}}$ are the corresponding eigenvalues energy of three
polaritons. It is worth noting that all three polaritons $b_{0}$, $b_{1}$, and
$b_{2}$ have been observed in intracavity EIT even with a hot ensemble
\cite{Wu1}. In the experiment \cite{Wu1}, a three-peak cavity transmission
spectrum, a very narrow peak and two broad peaks, can be clearly seen by
scanning the external field over a large frequency range. The dark-state
polariton $b_{0}$ (with the eigenvalue energy $E_{0}=0$) \cite{Fleisch}
corresponds to the narrow peak in the middle of transmission spectrum, and the
bright polaritons $b_{1}$ and $b_{2}$ correspond to two broad side peaks. The
intervals between resonant frequencies of central narrow peak and two broad
side peaks are determined by $E_{1}=\left\vert E_{2}\right\vert =\sqrt
{Ng^{2}+\Omega^{2}}$, the eigenvalues energy of the bright polaritons.

The blockade interaction via Rydberg level \textquotedblleft
hopping\textquotedblright\ is described by $H_{2}=\chi_{ij}%
{\displaystyle\sum\nolimits_{i>j}}
\left\vert r_{2}\right\rangle _{i}\left\vert r_{2}\right\rangle _{j}%
(\left\langle r_{1}\right\vert \left\langle r_{3}\right\vert +\left\langle
r_{3}\right\vert \left\langle r_{1}\right\vert )+H.c.$ \cite{Lukin1}, where
$\chi_{ij}\sim\wp_{r_{2}r_{1}}\wp_{r_{2}r_{3}}/r_{ij}^{3}$, $\wp_{r_{2}r_{1}}$
($\wp_{r_{2}r_{3}}$) is the dipole matrix element for the corresponding
transition and $r_{ij}$ is the distance between the two atoms. The level
states $\left\vert r_{1}\right\rangle $ and $\left\vert r_{3}\right\rangle $
are two sublevels of different parity. In general this interaction does not
affect the singly excited Rydberg state $\left\vert r_{2}\right\rangle $ but
leads to a splitting of the levels when two or more atoms are excited to the
state $\left\vert r_{2}\right\rangle $. The manifold of doubly excited states
of atomic ensemble trapped in a finite volume $V$ has an energy gap of order
$\bar{\chi}=\wp_{r_{2}r_{1}}\wp_{r_{2}r_{3}}/V$ \cite{Lukin1}. Using the
expression of $C_{r_{2}}$ in the polariton bases: $C_{r_{2}}=\cos\theta
(b_{1}+b_{2})/\sqrt{2}-\sin\theta b_{0}$, the Hamiltonian $H_{2}$ is then
given by
\begin{align}
H_{2}  &  =\bar{\chi}\sin^{2}\theta C_{r_{1}}^{\dagger}C_{r_{3}}^{\dagger
}b_{0}b_{0}+\frac{\bar{\chi}\cos^{2}\theta}{2}C_{r_{1}}^{\dagger}C_{r_{3}%
}^{\dagger}b_{1}b_{1}\nonumber\\
&  +\frac{\bar{\chi}\cos^{2}\theta}{2}C_{r_{1}}^{\dagger}C_{r_{3}}^{\dagger
}b_{2}b_{2}-\frac{\bar{\chi}\sin\theta\cos\theta}{\sqrt{2}}C_{r_{1}}^{\dagger
}C_{r_{3}}^{\dagger}b_{1}b_{0}\nonumber\\
&  -\frac{\bar{\chi}\sin\theta\cos\theta}{\sqrt{2}}C_{r_{1}}^{\dagger}%
C_{r_{3}}^{\dagger}b_{2}b_{0}+\frac{\bar{\chi}\cos^{2}\theta}{2}C_{r_{1}%
}^{\dagger}C_{r_{3}}^{\dagger}b_{2}b_{1}+H.c..
\end{align}
Equation (2) describes the nonlinear processes of the four-mode mixing
\cite{Walls}.

Now we consider the external fields interact with cavity mode through two
input ports $\alpha_{in}$, $\beta_{in}$, and two output ports $\alpha_{out}$,
$\beta_{out}$ with\ the interaction Hamiltonian \cite{Walls} $H_{3}=%
{\displaystyle\sum\nolimits_{\Theta=\alpha,\beta}}
[i%
{\displaystyle\int\nolimits_{-\infty}^{+\infty}}
d\omega\sqrt{\frac{\kappa}{2\pi}}\Theta^{\dagger}(\omega)ae^{i\Delta(\omega
)t}+H.c.]$, where $\Delta(\omega)=\omega-\omega_{cav}$ is the frequency
detuning of the external field from cavity mode, $\kappa$ is the cavity decay
rate, and $\Theta(\omega)$ with the standard relation $[\Theta(\omega
),\Theta^{\dagger}(\omega^{^{\prime}})]=\delta(\omega-\omega^{^{\prime}})$
denotes the one-dimensional free-space mode. We express the Hamiltonian
$H_{in-out}$ in the polariton bases: $H_{3}=%
{\displaystyle\sum\nolimits_{\Theta=\alpha,\beta}}
i%
{\displaystyle\int\nolimits_{-\infty}^{+\infty}}
d\omega\Theta^{\dagger}(\omega)[\sqrt{\frac{K_{0}}{2\pi}}b_{0}e^{i\Delta
(\omega)t}+\sqrt{\frac{K_{1}}{2\pi}}b_{1}e^{i\Delta(\omega)t}+\sqrt
{\frac{K_{2}}{2\pi}}b_{2}e^{i\Delta(\omega)t}]+H.c.$, with $K_{0}=\cos
^{2}\theta\kappa$, and $K_{1}=K_{2}=\sin^{2}\theta\kappa$. We assume that a
weak continuous-wave (cw) coherent field with the frequency $\omega$
resonantly drives the cavity mode. Then the effective non-Hermitian
Hamiltonian for the external field is%

\begin{equation}
H_{3}^{^{\prime}}=%
{\displaystyle\sum\nolimits_{\Lambda=0,1,2}}
(\Omega_{\Lambda}b_{\Lambda}e^{i\Delta(\omega)t}+H.c.)-\frac{iK_{\Lambda}}%
{2}b_{\Lambda}^{\dagger}b_{\Lambda},
\end{equation}
where $\Omega_{\Lambda}=\sqrt{2K_{\Lambda}}\beta$ with $\beta$ being the field
amplitude of the weak coherent field in natural units \cite{Imamo}.

Based on above analysis, the dynamics of our model is govern by the full
Hamiltonian $H=H_{1}+H_{2}+H_{3}^{^{\prime}}$. The quantum-state evolution of
this system is decided by Schr\"{o}dinger's equation ($\hslash=1$)
$i\partial_{t}\Psi(t)=H\Psi(t)$. We define $H=H_{0}+H_{int}$, with
$H_{0}=H_{1}$ and $H_{int}=H_{2}+H_{3}^{^{\prime}}$. After performing the
unitary transformations $\Psi^{^{\prime}}(t)=e^{iH_{0}t}\Psi(t)$ and
$H_{int}^{^{\prime}}=e^{iH_{0}t}H_{int}e^{-iH_{0}t}$, we can obtain
$i\partial_{t}\Psi^{^{\prime}}(t)=H_{int}^{^{\prime}}\Psi^{^{\prime}}(t)$
with
\begin{align}
H_{int}^{^{\prime}}  &  =[\bar{\chi}\sin^{2}\theta C_{r_{1}}^{\dagger}%
C_{r_{3}}^{\dagger}b_{0}b_{0}+\frac{\bar{\chi}\cos^{2}\theta}{2}C_{r_{1}%
}^{\dagger}C_{r_{3}}^{\dagger}b_{1}b_{1}e^{-i2E_{1}t}\nonumber\\
&  +\frac{\bar{\chi}\cos^{2}\theta}{2}C_{r_{1}}^{\dagger}C_{r_{3}}^{\dagger
}b_{2}b_{2}e^{-i2E_{2}t}-\frac{\bar{\chi}\sin\theta\cos\theta}{\sqrt{2}%
}C_{r_{1}}^{\dagger}C_{r_{3}}^{\dagger}\nonumber\\
&  \otimes b_{1}b_{0}e^{-iE_{1}t}-\frac{\bar{\chi}\sin\theta\cos\theta}%
{\sqrt{2}}C_{r_{1}}^{\dagger}C_{r_{3}}^{\dagger}b_{2}b_{0}e^{-iE_{2}%
t}\nonumber\\
&  +\frac{\bar{\chi}\cos^{2}\theta}{2}C_{r_{1}}^{\dagger}C_{r_{3}}^{\dagger
}b_{2}b_{1}e^{-i(E_{1}+E_{2})t}+H.c.]\nonumber\\
&  +%
{\displaystyle\sum\nolimits_{\Lambda=0,1,2}}
\{\Omega_{\Lambda}b_{\Lambda}e^{i[\Delta(\omega)-E_{\Lambda}]t}%
+H.c.\}\nonumber\\
&  -%
{\displaystyle\sum\nolimits_{\Lambda=0,1,2}}
\frac{iK_{\Lambda}}{2}b_{\Lambda}^{\dagger}b_{\Lambda}.
\end{align}

When $\bar{\chi}=0$, Equation (4) reduces to
\begin{align}
H_{int}^{^{\prime\prime}}  &  =%
{\displaystyle\sum\nolimits_{\Lambda=0,1,2}}
\{\Omega_{\Lambda}b_{\Lambda}e^{i[\Delta(\omega)-E_{\Lambda}]t}%
+H.c.\}\nonumber\\
&  -%
{\displaystyle\sum\nolimits_{\Lambda=0,1,2}}
\frac{iK_{\Lambda}}{2}b_{\Lambda}^{\dagger}b_{\Lambda},
\end{align}
which describes the interactions of the conventional intracavity EIT
\cite{Lukin,Wang,Hernandez,Wu1}. From Eq. (5), one can clearly see the
resonant condition for the polariton $b_{\Lambda}$: $\Delta(\omega
)=E_{\Lambda}$, as shown in the experiment \cite{Wu1}.

\section{Photon blockade}

When $\bar{\chi}\neq0$, \ Equation (4) describes the interactions of
intracavity EIT with Rydberg blockade interaction. Assuming that the external
field resonantly drive the polariton $b_{0}^{\dagger}$, i.e., $\Delta
(\omega)=E_{0}=0$, and the eigenvalues energy of the bright polaritons are
large enough, i.e., $E_{1}=\left\vert E_{2}\right\vert \gg\frac{\bar{\chi}%
\cos^{2}\theta}{2},\frac{\bar{\chi}\sin\theta\cos\theta}{\sqrt{2}}%
,\Omega_{\Lambda}$, one can realize a rotating-wave approximation and
eliminate from Eq. (4) the terms that oscillate with high frequencies
\cite{Solano}, leading to
\begin{equation}
H_{int}^{^{^{\prime\prime\prime}}}=(\lambda C_{r_{1}}^{\dagger}C_{r_{3}%
}^{\dagger}b_{0}b_{0}+\Omega_{0}b_{0}+H.c.)-%
{\displaystyle\sum\nolimits_{\Lambda=0,1,2}}
\frac{iK_{\Lambda}}{2}b_{\Lambda}^{\dagger}b_{\Lambda}.
\end{equation}
here $\lambda=\bar{\chi}\sin^{2}\theta$. Since the polaritons $b_{1}$, $b_{2}$
are not excited in Eq. (6), we disregard the decay terms including the
polaritons $b_{1}$, $b_{2}$, and obtain an effective interaction Hamiltonian
\begin{equation}
H_{eff}=(\lambda C_{r_{1}}^{\dagger}C_{r_{3}}^{\dagger}b_{0}b_{0}+\Omega
_{0}b_{0}+H.c.)-\frac{iK_{0}}{2}b_{0}^{\dagger}b_{0}.
\end{equation}
We note that the effective Hamiltonian here is similar to that used in
original scheme for photon blockade \cite{Imamo}. A photon from the probe
field is injected to a cavity to form the first Rydberg dark-state polariton.
However, injection of a second polariton will be blocked, since the presence
of the second polariton in the cavity will require an additional frequency
shift $\lambda$, which can not be provided by the incoming photons. Only after
the first polariton leaves the cavity can a second photon be injected. The
strong interactions between the polaritons therefore cause the strong photon
blockade effect.

\section{Discussion and numerical simulations}

Before supporting the numerical calculations, we briefly discuss the features
of intracavity EIT with blockaded Rydberg ensemble. The dominant decoherence
of intracavity EIT system associated with the leakage through the mirrors is
specified by the effective cavity decay rate $K_{0}=\kappa\cos^{2}\theta$,
which is determined by not only $\kappa$ but also the added factor $\cos
^{2}\theta$. With a decrease of $\cos\theta$ by tuning the control field, the
effective cavity decay rate $K_{0}$ decreases. A second dissipative channel is
the spontaneous emission of Rydberg states, luckily the Rydberg states have
long coherence time and small decay rate $\gamma_{r}\approx2\pi$ kHz
\cite{Saffman}. In order to have the strong blockade effect, the nonlinear
strength $\lambda$ should be much larger than the rate at which the
decoherences occur \cite{Imamo}. With a decrease of $\cos\theta$, the
nonlinear strength $\lambda=\bar{\chi}\sin^{2}\theta=\bar{\chi}(1-\cos
^{2}\theta)$ increases, while the effective cavity decay rate $K_{0}$
decreases. Hence one could achieve the strong-nonlinearity condition, i.e.,
$\lambda\gg K_{0}$, $\gamma_{r}$, by decreasing $\cos\theta$.
\begin{figure}[ptb]
\includegraphics[width=3.1in]{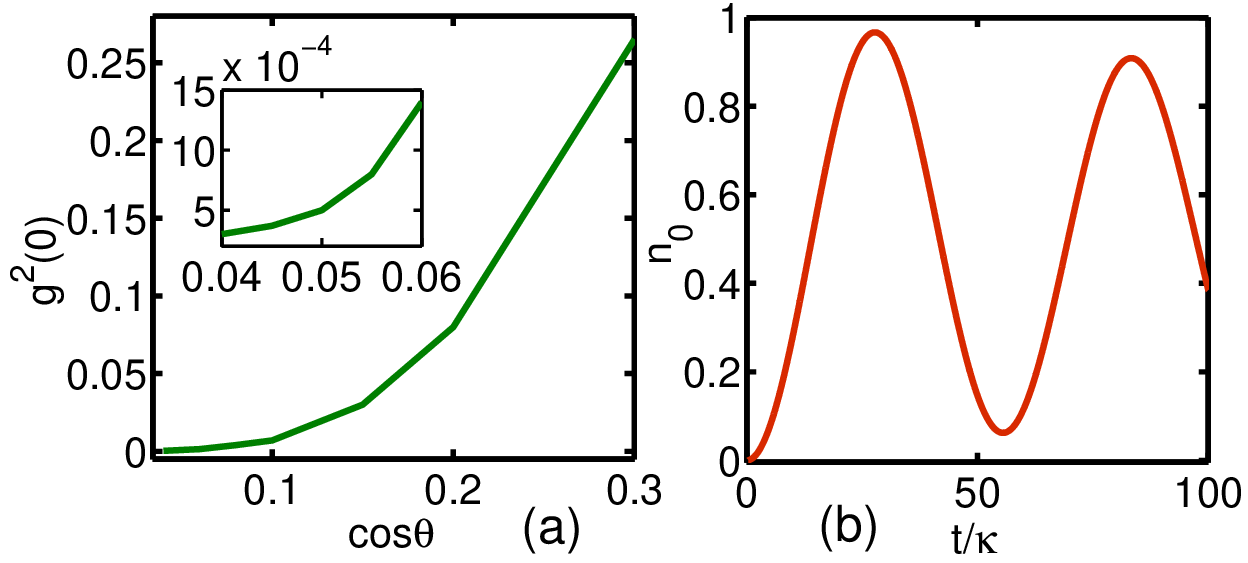}\newline\caption{(Color online) (a)
The second-order correlation function $g^{2}(0)$ as a function of $\cos\theta
$. (b) The number of dark-state polariton $n_{0}=\left\langle b_{0}^{\dagger
}b_{0}\right\rangle $ versus the normalized time $t/\kappa$, with $\cos
\theta=0.04$. Other common parameters: $\kappa=1$, $\gamma_{e}=\kappa$,
$\gamma_{r}=0.001\kappa$, $\bar{\chi}=2\kappa$, $g=3\kappa$, $\beta=1$ and
$N=600$.}%
\end{figure}

\begin{figure}[ptb]
\includegraphics[width=3.0in]{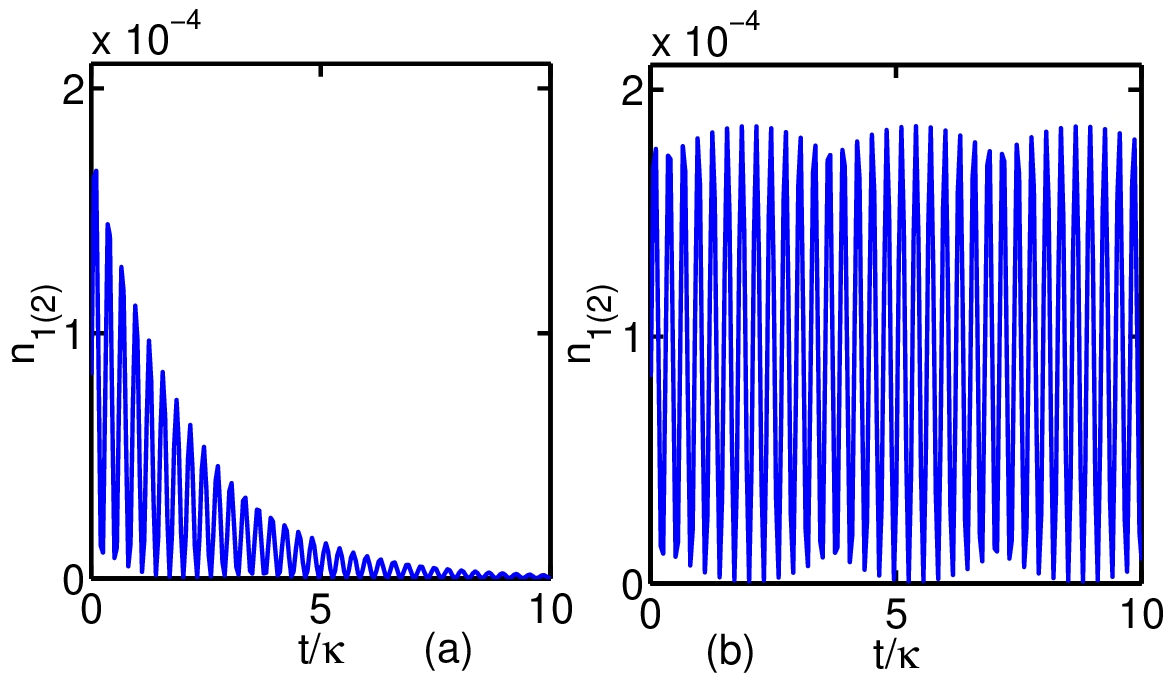}\newline\caption{(Color online) The
number of bright polariton $n_{1(2)}=\left\langle b_{1(2)}^{\dagger}%
b_{1(2)}\right\rangle $ versus the normalized time $t/\kappa$, for the cases
of$\ $(a) $\gamma_{e}=\kappa$ and (b) $\gamma_{e}=0$. Other common parameters:
$\kappa=1$, $\cos\theta=0.04$, $\gamma_{r}=0.001\kappa$, $\bar{\chi}=2\kappa$,
$g=3\kappa$, $\beta=1$ and $N=600$.}%
\end{figure}

To see the dynamics process of intracavity EIT with blockaded Rydberg
ensemble, we perform numerical simulations with the Hamiltonian (including the
atomic spontaneous emissions)
\begin{equation}
H_{num}=H-\frac{i\gamma_{r}}{2}b_{0}^{\dagger}b_{0}-\frac{i\gamma_{e}}{2}%
b_{1}^{\dagger}b_{1}-\frac{i\gamma_{e}}{2}b_{1}^{\dagger}b_{1},
\end{equation}
with $\gamma_{e}$ being the decay rate of excited state $\left\vert
e\right\rangle $. Our investigation relies on calculations of photon
correlations based on the normalized intensity correlation function
$g^{2}(\tau)$ \cite{111} for the dark-state polariton $b_{0}$. The incident
coherent classical field has $g_{in}^{2}(\tau)=1$, corresponding to a Poisson
distribution for photon number independent of time delay $\tau$. The ideal
photon blockade would achieve $g^{2}(0)=0$, in correspondence to the state of
a single photon. More generally, $g^{2}(0)<1$ represents a nonclassical effect
with the variance in photon number reduced below that of the incident field.
In Fig. 2 (a), we plot numerical simulations results of $g^{2}(0)$ as a
function of $\cos\theta$. From Fig. 2(a), we see that $g^{2}(0)$ decreases
with a decrease of $\cos\theta$. When $\cos\theta=0.04$, $g^{2}(0)$\ would be
in the order of $10^{-4}$. Figure 2(b) shows the number of dark-state
polariton $n_{0}=\left\langle b_{0}^{\dagger}b_{0}\right\rangle $ versus the
normalized time $t/\kappa$. We clearly see that the number of cavity
polaritons varies between $0$ and $1$, but never exceeds unity. These
numerical simulations results show that the system could behave very strong
photon blockade effect. Figure 3 (a) and (b) show the number of bright
polariton $n_{1(2)}=\left\langle b_{1(2)}^{\dagger}b_{1(2)}\right\rangle $ for
the cases of $\gamma_{e}=\kappa$ and $\gamma_{e}=0$. Under certain condition,
the number of bright polariton $n_{1(2)}$ could be in the order of $10^{-4}$,
thus the excitation of bright polaritons can be neglected when the external
field resonantly drive the dark-state polariton $b_{0}$.

We note that Souza et al. \cite{Souza} have theoretically studied photon
blockade and antiblockade effects with the single-atom ($N=1$) EIT in an
optical cavity. In the approach of Ref. \cite{Souza}, one of the bright
polaritons is excited, while the excitation of the dark-state polariton can be
neglected. In contrast to this approach, we choose to drive the dark-state
polariton (the excitation of bright polaritons can be neglected). Since the
dark-state polariton has no contribution from the excited state $\left\vert
e\right\rangle $, there is longer coherence time in our scheme than that in
Ref. \cite{Souza}. To excite one of three polaritons independently, large
intervals between the resonant frequencies of three polaritons are required.
Obviously, with an ensemble of $N$ ($N\gg1$) atoms, our scheme could satisfy
this requirement more easily. We also note that Peyronel et al.
\cite{Peyronel} have experimentally demonstrated a quantum nonlinear
absorption filter with EIT in a blockaded Rydberg ensemble with $g^{2}(0)$ in
the order of $10^{-2}$. In contrast to this protocol, our proposed scheme
combines the advantage of variable cavity lifetimes with intracavity EIT and
the power of strong interactions among Rydberg atoms, thereby achieving strong
photon blockade effect with smaller value of $g^{2}(0)$ in the order of
$10^{-4}$.

Now we address the experiment feasibility of the proposed scheme. An ensemble
of $600$ cold atoms is trapped in an optical cavity \cite{Colombe}. For the
high Rydberg states $n\geq100$, the blockade interaction strength over the
atomic sample is $\bar{\chi}/2\pi\approx100$ MHz \cite{Saffman}. For the
relevant cavity parameters, $(\kappa,\gamma_{e})/2\pi\approx(53,3)$ MHz, and
$g/2\pi\approx200$MHz \cite{Colombe}. With the choice $\cos\theta=0.04$ and
$\beta=7$, we have $E_{1}=\left\vert E_{2}\right\vert \approx2\pi\times4898$
MHz, $\frac{\bar{\chi}\cos^{2}\theta}{2}\approx2\pi\times0.08$ MHz,
$\frac{\bar{\chi}\sin\theta\cos\theta}{\sqrt{2}}\approx2\pi\times2.82$ MHz,
$\Omega_{0}=\sqrt{2K_{0}}\beta\approx2\pi\times2.8$ MHz, $\Omega_{1(2)}%
=\sqrt{2K_{1(2)}}\beta\approx2\pi\times70$ MHz, thus the condition
$E_{1}=\left\vert E_{2}\right\vert \gg\frac{\bar{\chi}\cos^{2}\theta}{2}%
,\frac{\bar{\chi}\sin\theta\cos\theta}{\sqrt{2}},\Omega_{\Lambda}$ is well
satisfied. Then $\lambda=\bar{\chi}\sin^{2}\theta\approx2\pi\times99.8$ MHz is
three orders of magnitude larger than both the effective cavity decay rate
$K_{0}=\cos^{2}\theta\kappa\approx$ $2\pi\times0.09$ MHz and Rydberg atomic
decay rate $\gamma_{r}$. Choosing even smaller values of $\cos\theta$ would no
longer make sense because then the decay of Rydberg states dominates
decoherence of intracavity EIT system.

\section{Conclusion}

In summary, we have presented a scheme for strong photon blockade with
intracavity EIT in a cold ensemble of strongly interacting Rydberg atoms, by
combining the advantage of the variable cavity lifetimes with intracavity EIT
and strongly interacting Rydberg dark-state polaritons. Our scheme for strong
photon blockade suggests intriguing prospects for quantum simulation of the
rich physics promised by strongly-correlated quantum systems in cavity QED.
For example, with an array of coupled cavities, if each cavity consists of an
atomic ensemble to effectively generate strong photon blockade interaction,
the interplay of photon blockade interaction and tunnelling coupling between
neighboring cavities will lead to interesting many-body physics \cite{Michael
J}.

\textbf{Acknowledgments: }This work was supported by the National Natural
Sciences Foundation of China (Grants No. 11204080, No. 11274112, No. 91321101,
and No. 61275215), the Shanghai Municipal Natural Science Foundation (Grant
No. 13ZR1411700), and the Fundamental Research Funds for the Central
Universities (Grants No. WM1313003).

\end{document}